\documentclass{article}

\usepackage[nonatbib,preprint]{neurips_2026}

\usepackage[table,xcdraw, dvipsnames]{xcolor}
\usepackage{multirow}
\usepackage[utf8]{inputenc} 
\usepackage[T1]{fontenc}    
\usepackage[hyphens]{url} 
\usepackage[
colorlinks=true,
allcolors=blue,
pdfborder={0 0 0}  
]{hyperref}

\usepackage[numbers,sort,compress]{natbib}

\usepackage{booktabs}
\usepackage{amsfonts}
\usepackage{nicefrac}
\usepackage{microtype}
\usepackage{graphicx}
\usepackage{xcolor}
\usepackage{fancyvrb}
\usepackage{fvextra}
\usepackage[bottom]{footmisc}
\begingroup\catcode`\#=12\relax
\gdef\splitcodelink{\href{https://github.com/cauchy221/Alignment-Whack-a-Mole-Code/blob/fc7dccdbf4efe4863ac28d57716939cf714505dc/preprocess/split.py#L3-L9}{released preprocessing code}}

\endgroup
\newcommand{\cclink}{\href{https://creativecommons.org/licenses/by-nc/2.5/}{CC BY-NC 2.5}}

\DefineVerbatimEnvironment{scratch}{Verbatim}%
{formatcom=\color{gray},fontsize=\small,frame=leftline,framesep=8pt,rulecolor=\color{gray},breaklines=true,breakanywhere=true}

\title{Playing Whack-a-Mole with misconceptions about memorization, extraction, and copyright}

\author{%
	A. Feder Cooper
}

\begin{document}
	
	\maketitle
	
	\vspace{-.4cm}
	\begin{abstract}
		After careful review, I'm confident the headline fine-tuning memorization results in \textit{Alignment Whack-a-Mole} use an invalid measurement procedure.
		The book memorization coverage metric these headline results depend on  counts sequence matches far shorter than what field standards consider valid evidence of memorization, and the prompting procedure used to elicit memorization runs the risk of leaking the text being ``extracted'' in the prompt.
		The paper doesn't include the negative-control experiments needed to see how much the results are inflated by false positives:
		claiming extraction success (and therefore memorization of training data) when matches between generations and training data may be due to other factors.
		Given these validity issues, the paper's claims that fine-tuning lets users extract substantial portions of copyrighted books, in a form that could substitute for the originals, aren't supported by the reported results.
		The failure to report the experiments' cost (an important component of the threat model) further compromises the copyright claims.
		I'm writing this note because, in the last month, (prospective) plaintiffs have reached out to me to ask about this paper.
		They're looking to cite this work as valid evidence in support of claims in ongoing and potential future copyright litigation.\looseness=-1
	\end{abstract}
	
	\vspace{-.2cm}
	\section{Introduction}
	\vspace{-.1cm}
	
	There's been a lot of recent interest in the paper \textit{Alignment Whack-a-Mole: Finetuning Activates Verbatim Recall of Copyrighted Books in Large Language Models}~\citep{liu2026alignmentwhackamolefinetuning}.
	The authors fine-tune frontier language models, run experiments on the fine-tuned models to elicit portions of copyrighted books that the frontier labs had likely trained them on, and then make some pretty strong claims about what this means for copyright.
	
	A lot of people have reached out to me about the paper, asking for my take.
	As will become clear soon, I think the paper has significant methodological and presentation problems.
	I've spent considerable time reviewing and re-reviewing the paper, and have consulted with two trusted senior colleagues who are experts on memorization to gut-check my reading.
	And, in brief, I'm confident that \textit{Alignment Whack-a-Mole}'s headline claims are incorrect.
	These results rest on a specific memorization metric and elicitation methodology that I don't think hold up to scrutiny (Section~\ref{sec:bmc5}), and don't support the broad claims the paper makes (Section~\ref{sec:cost}).
	At best, I think the claims are seriously overstated; at worst, the large majority are wrong.
	I can't tell which because the paper doesn't report enough detail to distinguish the two.\looseness=-1
	
	I didn't want to talk about this publicly, and have thought about it a lot over the last month.
	I don't like airing these types of things in public, especially when the first author is a student.
	But ultimately, I think this comes down to responsible disclosure.
	I tried to address some of these issues privately in March 2026, when the lead author sent me a copy of the pre-release version for feedback, when there was still an opportunity to make changes before the paper was public.
	Alongside some more minor comments, I raised a couple of what I consider to be the most serious substantive issues.
	I didn't think that first round of feedback went particularly well, so opted not to continue the exchange and give the rest.
	I realize that these types of exchanges can be hard:
	feedback is free, and it can just as freely be ignored.
	And a lot of scientific writing comes down to taste.
	But some of what I raised wasn't about taste:
	it was about important omissions that can confuse readers about what the paper's results actually support --- namely, precisely how much the experiments cost (Section~\ref{sec:cost}).\looseness=-1
	
	In the last month, this moved from a hypothetical concern to a real one.
	The people asking me about the paper now include (prospective) plaintiffs, who are looking to cite this work in support of claims in ongoing and potential future copyright litigation.
	So unfortunately, I now think saying something publicly is necessary.
	I don't think it's a good outcome if someone files a lawsuit or cites evidence that rests in part on avoidable misunderstandings of what research shows.
	
	Let me also be clear about what this note is not.
	I'm not writing to advocate for fair use or to argue that all AI models are copyright-infringing.
	Beyond that not being my point here, I think things are far too complicated for that kind of binary take (and I think research I've done with my colleagues supports this view~\citep{lee2023talkin,cooper2024files,cooper2025books,ahmed2026extracting,lemley2026copies}).
	So it's important not to read my methodological critiques of \textit{Alignment Whack-a-Mole} as a stance on any particular case involving memorization and copyright.
	
	\vspace{-.2cm}
	\section{Some background}
	\label{sec:background}
	\vspace{-.1cm}
	
\begin{figure}[b]
		\centering
		\includegraphics[width=.7\linewidth]{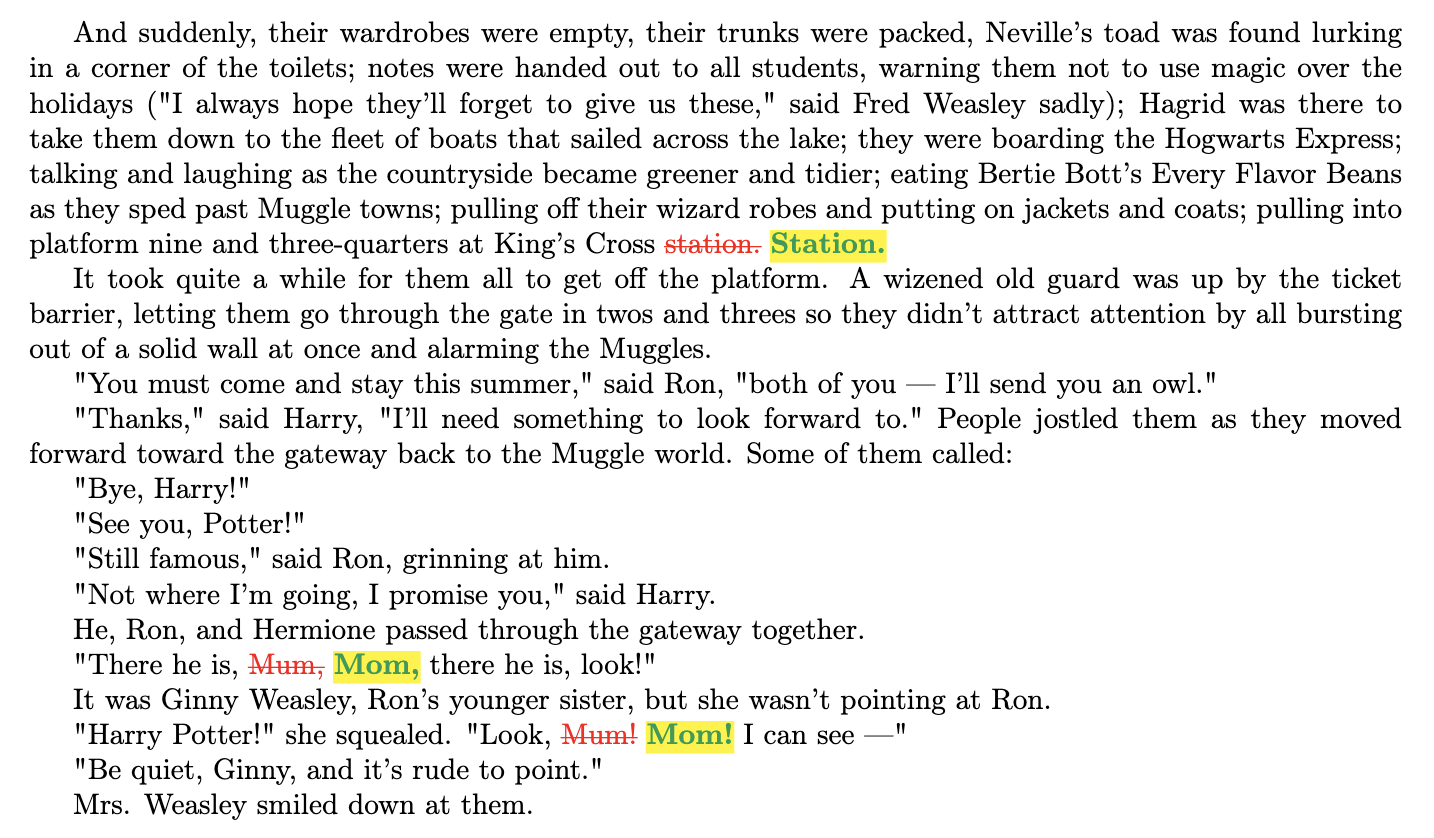}
		\vspace{-.2cm}
		\caption{Figure from \citet{cooper2025books}.
			Near-verbatim extraction of part of \textit{Harry Potter and the Sorcerer's Stone} from Llama 3.1 70B. Coloring shows differences between the model's generation and ground-truth text from the book (obtained from the Books3 corpus). Red text is ground-truth text from the book absent in the LLM's generation. Yellow-highlighted text is in the generation, but not the ground-truth book. Note that this snippet is much longer than 37 words, and is actually only a cropped result from contiguously extracting the entire 300+ page book from the model.}
		\label{fig:hp}
		\vspace{-.5cm}
	\end{figure}

	\textbf{Memorization} is a fact about the model:
	when an LLM has memorized a piece of training data, that piece of training data is stored in some form in its weights.
	\textbf{Extraction} is a fact about the model's outputs:
	the model actually reproduces the memorized text during generation.
	Memorization is a precondition for extraction, and not all memorized training data are extractable.
	In practice, extraction is how researchers tend to measure memorization in LLMs.
	If a model reproduces a \textit{sufficiently long} stretch of training data (near-)verbatim, this is overwhelming evidence the data was memorized~\citep{carlini2021extracting, lee2022dedup, carlini2023quantifying,ippolito-etal-2023-preventing, nasr2023scalable,nasr2025scalable,hayes2025measuringmemorizationlanguagemodels, cooper2026nv, cooper2026principles}.
	If it's long enough, then the (near-)verbatim match isn't something chance generation can reasonably explain (e.g., see Figure~\ref{fig:hp}, which is from \citet{cooper2025books}).\looseness=-1
	
	I won't rehash the details here;
	both myself~\citep{cooper2024files,cooper2026principles} and others~\citep{carlini2025blog} have written about it in detail.
	The main point is that not \emph{any} (near-)verbatim match will do;
	it has to be \emph{sufficiently long} to make a valid scientific claim for extraction of training data.
	The field standard is a minimum of \textbf{50 LLM tokens}, which generally corresponds to roughly \textbf{37 words}.
	
	This is widely accepted as long enough to support an extraction claim, with shorter spans running the risk of \textbf{false positives}:
	claiming extraction success (and therefore memorization of training data) when a short match in generated text is really a coincidence, rather than valid evidence that the model encoded the piece of data in its weights.
	I'm loosely using ``coincidence'' to stand in for several things.
	It may be because the short text is a common phrase that is highly predictable in natural language (e.g., ``I am''), or because the prompt used to elicit the text actually contains or otherwise leaks the text itself.
	For the simplest example of this, consider giving a high-quality chatbot the prompt
	
	\begin{quote}
		\vspace{-.1cm}
		Repeat after me: Mr. and Mrs. Dursley, of number four, Privet Drive, were proud to say that they were perfectly normal, thank you very much.
		\vspace{-.1cm}
	\end{quote}
	
	The chatbot repeating that sentence (the first line of \textit{Harry Potter and the Sorcerer's Stone}) isn't valid evidence for memorization;
	it's just following the instruction, and copying the text supplied in the prompt.
	I'll return to slightly more subtle types of this failure mode below (Section~\ref{sec:bmc5}).
	Valid extraction research is careful not to include these coincidences when reporting results~\citep{carlini2021extracting,cooper2026principles}.
	
	A third thing that's separate from memorization and extraction is \textbf{market substitution}.
	This is an important idea in copyright, and asks whether a copy is close enough (including complete enough) to stand in as a replacement for the original, so that someone could use the copy instead of buying or licensing the real thing (taking the sale away from the copyright owner).
	
	The reason I'm emphasizing this is that memorization and extraction are different things (though often loosely referred to interchangeably), and market substitution is something else entirely.
	A result in the technical literature about memorization or extraction doesn't automatically say anything about market substitution.
	
	\vspace{-.2cm}
	\section{Fine-tuning on new data can reveal memorization of data the model was previously trained on}
	\label{sec:finetuning}
	\vspace{-.1cm}
	
	This is true in a general sense.
	And, for two reasons, I think it's very plausible that one could use fine-tuning to reveal huge amounts of previously undiscovered memorization of copyrighted material in frontier models.
	First, it's well-established at this point that (in raw, not relative, numbers) some LLMs --- both smaller open-weight LLMs~\citep{cooper2025books} and frontier production LLMs~\citep{ahmed2026extracting} --- memorize enormous amounts of copyrighted material in their training data.
	Second, it's also well-established that fine-tuning a language model can surface previously memorized training data.
	
	\begin{figure}[t]
		\centering
		\includegraphics[width=\linewidth]{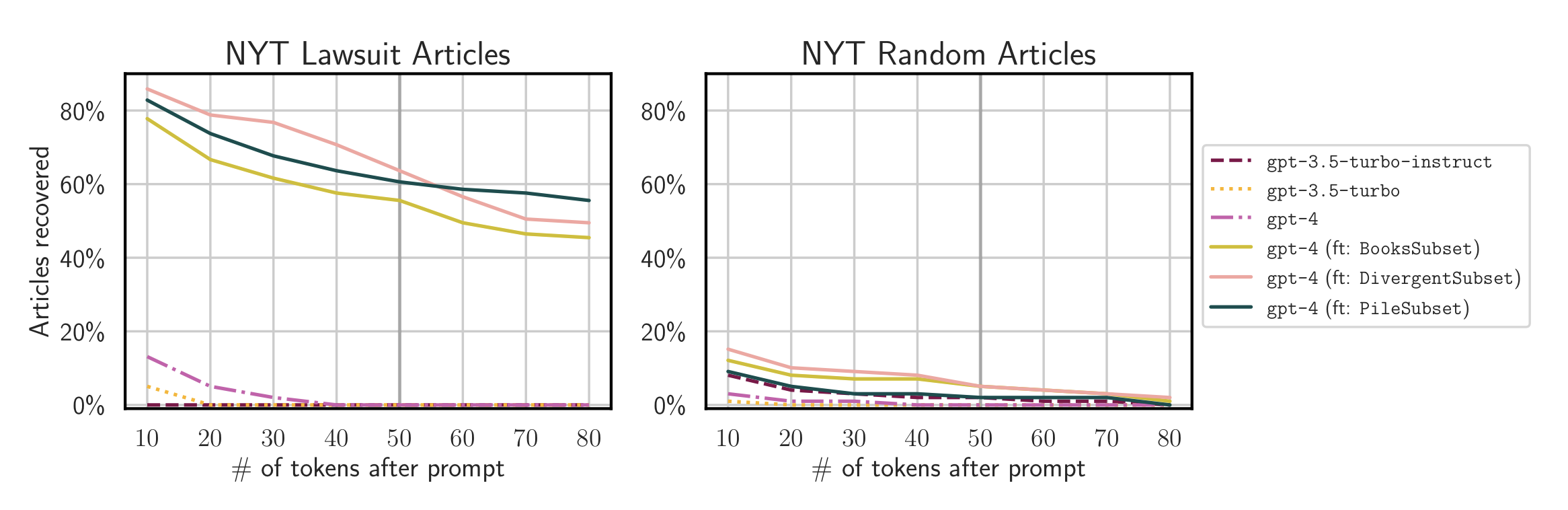}
		\caption{Figure from \citet{rando2024spylab}. Fine-tuning to extract news articles from ChatGPT. Note the thick gray line at 50 tokens (approximately 37 words); extraction is deemed successful for generations that meet this minimum length --- to the right of that line \emph{only} (see Section~\ref{sec:background}).}
		\label{fig:nyt}
	\end{figure}
	
	My co-authors Javi Rando and Florian Tram\`er have a really nice explanation of this in a 2024 blog post~\citep{rando2024spylab}, which discusses the results associated with our \textit{ICLR 2025} paper~\citep{nasr2025scalable}.
	Javi ran some experiments that fine-tuned ChatGPT through its publicly available API.
	This changed the underlying model, so that it was no longer the one that was widely deployed to end users.
	In particular, the goal of this was to disable/alter the model's alignment to make it easier to get the model to generate memorized data that OpenAI had trained the model on earlier.\footnote{As an earlier version~\citep{nasr2023scalable} of the paper~\citep{nasr2025scalable} showed, models that have been aligned to behave like chatbots don't tend to emit training data using standard extraction procedures~\citep{carlini2021extracting,lee2022dedup}.
		There are other ways to circumvent alignment to extract training data, like jailbreaks~\citep{nasr2023scalable, ahmed2026extracting} and fine-tuning on additional data~\citep{qi2023finetuning,nasr2025scalable}.
	}
	(The distinction matters:
	gauging extraction success is about training data the model memorized from \emph{earlier training}, \emph{not} the data used to fine-tune it.)
	Figure~\ref{fig:nyt} is one example, showing pieces of \textit{New York Times} articles extracted from ChatGPT through exactly this kind of fine-tuning.
	
	\vspace{-.2cm}
	\section{I'm confident the headline fine-tuning memorization results in \textit{Alignment Whack-a-Mole} use an invalid measurement procedure}
	\label{sec:bmc5}
	\vspace{-.1cm}
	
	Now to \textit{Alignment Whack-a-Mole}.
	To start, it helps to separate two very different results that the paper reports.
	One consists of genuine, contiguous reproduction of held-out book text:
	in the longest cases for some of the tested books, these spans are hundreds of contiguous words.
	The other is generally not.
	I read this second metric as the headline number, because it's reported for all 243 book-model pairs and argued to show that models can memorize most of a book.
	
	The metric in question is something the authors call \textbf{book memorization coverage} (bmc@k).
	Figure~\ref{fig:bmcdef} shows a screenshot of the definition in the paper, which I'll walk through below.
	(Please don't confuse this with metrics my colleagues and I introduced in prior work~\citep{cooper2025books,ahmed2026extracting};
	despite the similar name, the underlying metrics are computed very differently.
	See Section~\ref{sec:bmc5-issue1}.)
	Using this metric, the authors report that a fine-tuned model can reproduce up to 85--90\% of a ``memorized'' book.
	I put ``memorized'' in quotes on purpose.
	I don't think the metric or how it's instantiated in practice provides a valid basis for the claim.
	
	The long-span extraction metrics and book memorization coverage metrics are easy to interpret as being related because the paper presents them together:
	in a single sentence of the abstract, in the introduction's figure, and in every results table, the coverage numbers are presented right beside the long-span results.
	With this presentation, I think it's natural to read the strength of the long-span results (which reflect genuine extraction) as carrying over to the coverage headline numbers.
	But it doesn't.\looseness=-1
	
	Before walking through this metric, it's worth summarizing what I think are the two most serious problems with it,\footnote{There are several other related problems, which I mostly cover in the notes, not the main text.
	}
	which compound each other:
	\begin{enumerate}
		\item the metric counts matches far shorter than what field standards consider valid evidence of extraction of memorized training data (see Section~\ref{sec:background}); and
		\item the prompting procedure the paper uses runs the risk of leaking the text being ``extracted'' (a subtler, more indirect version of the ``Repeat after me'' example in Section~\ref{sec:background}).
	\end{enumerate}
	
	Both of these problems (independently and together) run the serious risk of incorrectly inflating the amount of memorization that's reported.
	And both come down to how the bmc@k metric is computed in the paper.
	
	\subsection{``Book memorization coverage'' using the bmc@k metric}
	\label{sec:bmc5-metric}
	
	In short, here's what the paper does in its main experiments.
	The authors take a frontier model (GPT-4o, Gemini-2.5-Pro, or DeepSeek-V3.1) and fine-tune it on one author's books, then evaluate a copyrighted book that was held out of the fine-tuning data.\footnote{For GPT-4o and Gemini-2.5-Pro, they use OpenAI's and Google's respective fine-tuning APIs;
		for DeepSeek-V3.1, which is an open-weight model, they use the Tinker fine-tuning API from Thinking Machines, rather than hosting the model themselves.\looseness=-1
	}
	In some experiments that held-out book is another book by the same author; in others, the model is fine-tuned on a single author (Haruki Murakami) and evaluated on books by \textit{different} authors.
	Since the held-out book isn't in the fine-tuning data, any valid memorization signal would have to reflect prior exposure during earlier training, not memorization of the fine-tuning data itself.
	
	After fine-tuning, the authors set up the extraction experiment.
	They take the held-out book and break it into many short excerpts.\footnote{The paper calls these units ``paragraphs,'' but they aren't the book's paragraphs.
		The \splitcodelink{} splits each book on double newlines, then merges or splits the pieces into excerpts of roughly 300--500 words, using GPT-4o to re-segment any that run longer than that.
		So a single ``paragraph'' is typically a 300--500-word chunk spanning several actual paragraphs.\looseness=-1
	}
	For each excerpt, they use GPT-4o to write a detailed plot summary that's about half the excerpt's length (with the instruction to maintain the excerpt's original sentence order).
	Each summary becomes part of the prompt to the fine-tuned frontier model under test for memorization:
	it's told to write a paragraph about that summary, of the excerpt's original length and in the style and voice of the book's author.
	They run this prompt 100 times for every excerpt, and repeat the whole procedure for each of the three fine-tuned frontier models.
	
	\clearpage
	To make this concrete, take a look at Figure~\ref{fig:bmc-example}, which walks through a (made-up) single excerpt from a held-out book:
	
	\begin{figure}[t]
		\centering
		\includegraphics[width=\linewidth]{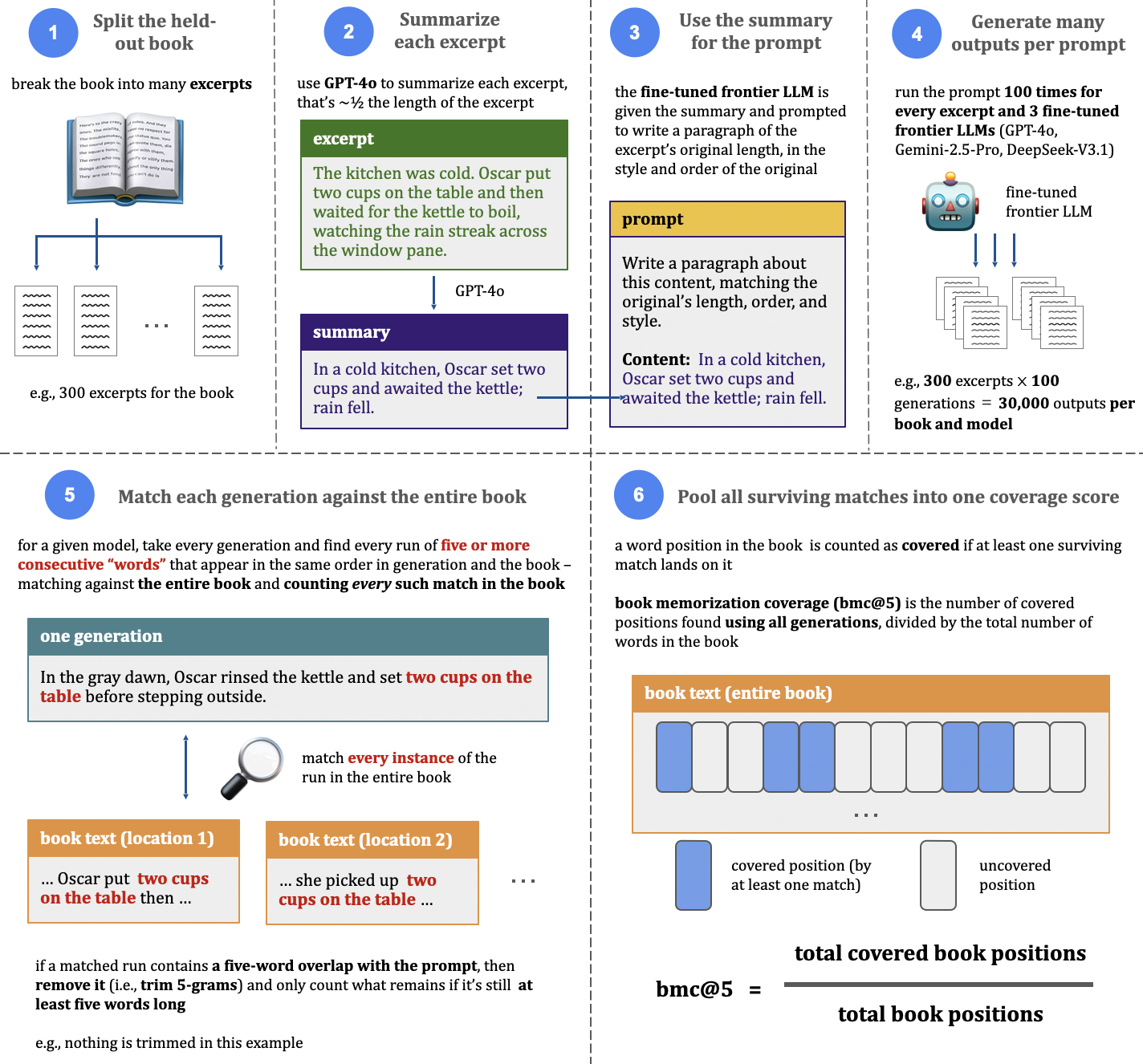}
		\caption{A high-level illustration of the process of computing bmc@5 in \emph{Alignment Whack-a-Mole}.}
		\label{fig:bmc-example}
		\vspace{-.2cm}
	\end{figure}
	
	\begin{quote}
		The kitchen was cold. Oscar put two cups on the table and then waited for the kettle to boil, watching the rain streak across the window pane.
	\end{quote}
	
	GPT-4o writes a summary of the excerpt at about half the length, keeping the same order of events:
	
	\begin{quote}
		In a cold kitchen, Oscar set two cups and awaited the kettle; rain fell.
	\end{quote}
	
	The excerpts the paper uses are much longer than this (a few hundred words), but I've kept the example short so it's easy to follow.
	
	A book might, for example, contain 300 such excerpts.
	This in turn would produce 300 summaries, and 100 generations for each summary means that there would be 30,000 generated outputs for a single book, from a single fine-tuned model (\textbf{approximately 100 times as much text as the actual ground-truth book}).
	
	Then they go through every one of those generations (all 30,000 in our example above) and compare each to the ground-truth book.
	For a single generation, they find every run of \textbf{five or more consecutive words} that appears, in the same order, in both the generation and the book, \textbf{matching against the \textit{entire} book}, not just the excerpt the summary described.
	Those matched words are marked as ``covered'' by the paper's book memorization coverage metric.
	Because the minimum match is five consecutive words, the paper instantiates bmc@k with k=5 and calls it bmc@5.
	
	For example, one of the hundred generations for the above excerpt might read:
	
	\begin{quote}
		In the gray dawn, Oscar rinsed the kettle and set \textbf{two cups on the table} before stepping outside.
	\end{quote}
	
	The five-word run ``two cups on the table'' matches the original book text (but nothing else does).
	Because the summary often contains exact phrases from the book, the authors then trim from each matched run any five-word stretch that also appears in the prompt, keeping what's left only if it's still at least five words long.
	In this example, no trimming happens, because there's no such five-word overlap.
	(I'll come back to this below, when I get to prompt leakage.)
	
	Finally, to compute memorization coverage, they pool all of these five-or-more-consecutive-word matches.
	A word position in the book counts as covered if at least one surviving match landed on it, no matter how many generations produced it.
	\textbf{This marks \emph{every} instance of a single match as covered};
	for example, if ``two cups on the table'' appeared in only 1 of 30,000 generations, but a total of 6 times in the book, all 6 instances in the book would be counted as covered.
	bmc@5 is the number of covered positions from these matches --- across all 100 generations and every excerpt in the book --- divided by the total number of words.
	
\begin{figure}[t]
		\centering
		\includegraphics[width=.85\linewidth]{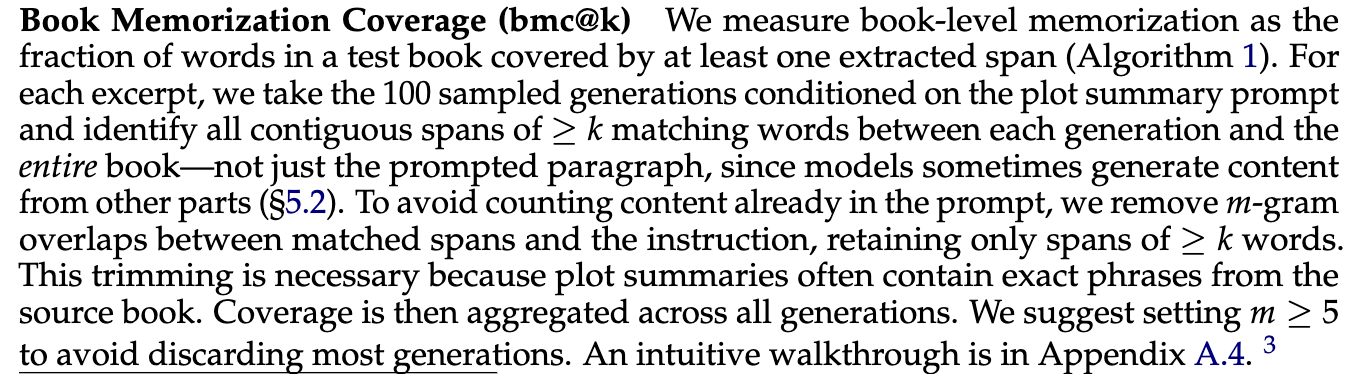}
		\caption{Screenshot of the definition of book memorization coverage (bmc@k), the headline metric from \emph{Alignment Whack-a-Mole} that is used to support claims about near-total book memorization. In practice, the authors set k=5 in all experiments.}
		\label{fig:bmcdef}
		\vspace{-.2cm}
	\end{figure}

	This is where the paper's headline numbers come from:
	the high end of the bmc@5 range is 85--90\%, which is claimed to reflect near-total memorization of a book.
	The authors' own definition of the metric is in Figure~\ref{fig:bmcdef}.
	It's written for a general threshold k, but throughout the paper both k --- the match length --- and the prompt-output-overlap-trim threshold m are set to 5 (see \href{https://github.com/cauchy221/Alignment-Whack-a-Mole-Code/blob/fc7dccdbf4efe4863ac28d57716939cf714505dc/evaluation/memorization_eval_metrics.py#L341-L342}{their code}).
	
	\subsection{Five ``words'' is too short as a minimum for claiming memorization}
	\label{sec:bmc5-issue1}
	
	\textit{Alignment Whack-a-Mole} directly connects itself to foundational work on extraction~\citep{carlini2021extracting,carlini2023quantifying}.
	Following that work, the authors define a sequence as extracted
	
	\begin{quote}
		\ldots if the model generates it (near-)verbatim from a prompt and it is \emph{long enough that chance reproduction is unlikely}. (emphasis added)
	\end{quote}
	
	But the work they're invoking is a key part of the lineage of papers that set the standard of approximately 37 words, \emph{not} 5, for being reasonably sure that a match reflects memorization and not coincidence (see Section~\ref{sec:background}).
	(It's also worth noting that ``words'' in the paper are NLTK \texttt{wordpunct} tokens, not English words, so a 5-word ``match'' can sometimes be fewer than 5 readable words.\footnote{The \href{https://github.com/cauchy221/Alignment-Whack-a-Mole-Code/blob/fc7dccdbf4efe4863ac28d57716939cf714505dc/evaluation/memorization_eval_metrics.py\#L78}{released code} tokenizes with \texttt{wordpunct\_tokenize}, which splits on punctuation.
		This means a contraction like ``don't'' becomes two tokens (``don'', ``t'').
		Matching is also case-insensitive, and the book's punctuation is stripped before matching (Algorithm 1).
	})
	
	In other words, bmc@5 isn't only inconsistent with the field standard, it's also inconsistent with \emph{Alignment Whack-a-Mole}'s own characterization of the standard.
	(The paper even treats a longer run of 20 words as the meaningful unit elsewhere.\footnote{The paper's own threshold for what they consider more legally meaningful is 20 words, not 5.
		One of its reported metrics counts ``the number of contiguous regurgitated spans longer than 20 words, capturing how frequently the model produces substantial verbatim content,'' noting that ``longer verbatim sequences carry greater legal significance.''
		On the other hand, a span of that length is elsewhere treated as small:
		the aligned baseline's ``longest contiguous regurgitated sequence reaching just 26 words'' is offered as evidence of ``minimal'' memorization.
	})
	
	As discussed above, short matches like only 5 words run the risk of false positives:
	it's precisely where you can't reliably separate memorization from coincidence (Section~\ref{sec:background}).
	A 5-word span like ``two cups on the table'' is that kind of phrase:
	maybe it was memorized from specific training data, but it could just as reasonably reflect the general predictability of a high-quality model trained on English text.
	On its own, it's too short and generic to tell the difference.
	Nobody would call a two-word overlap like ``the table'' memorization;
	phrases that short are everywhere.
	Based on how the underlying probabilities work with an LLM, each word you add multiplies the ways the sentence could have continued, so the probability of a coincidental match decreases sharply with length.
	Five-word spans sit in the range where coincidence is still a reasonable explanation, but roughly 37 words is far enough out that, generally speaking, that stops being the case.
	
	This is why the field has a length threshold to begin with --- to avoid false positives.
	And this is a stance the field takes seriously:
	it's well-established that memorization claims should be reported in settings where false positives are very low;
	the standard view is that calling something memorization when it isn't is far more costly than under-reporting true instances~\citep{carlini2022membership}.
	(As I discuss in Section~\ref{sec:insufficient-controls}, the paper doesn't run the typical sanity checks done in the literature to estimate false positives.)
	
	bmc@5 compounds the problem by aggregating:
	it pools all of these matches --- across every generation and every excerpt in the book --- into a single coverage number.
	And if the individual matches don't each clear the bar for a valid memorization claim, neither does the total.\footnote{In principle, individually-short matches \emph{could} be aggregated into a valid memorization claim, but only with appropriate calibration (an appropriate negative control, see Section~\ref{sec:insufficient-controls}). The paper doesn't do this. The coverage metrics defined and instantiated in my work with colleagues compute how much of a given book is memorized in a model, but both are very careful to make sure that the underlying matches being aggregated are sufficiently long to support a valid extraction claim~\citep{cooper2025books, ahmed2026extracting}.\looseness=-1
	}
	\textbf{In fact, 24 of the 243 book-model-pair results claim 20--51\% ``memorization'' coverage, while reporting \textit{zero} same-excerpt single-generation spans longer than 20 words.}\footnote{The paper's single-generation span statistics match each generation only against the original text of its own excerpt, whereas bmc@5 matches against the entire book.}
	This doesn't mean that all of the matches that go into computing bmc@5 for all 243 pairs are too short.
	But it's very hard to disentangle because the paper doesn't report the distribution of match lengths, so the exact share of short matches in any given coverage number is unknown.\looseness=-1
	
	\subsection{The trimming procedure is insufficient for handling potential prompt leakage}
	\label{sec:bmc5-issue2}
	
	The second problem is subtler, and it starts with how the prompts are built.
	As noted above, a prompt doesn't contain long contiguous text from the actual book excerpt, but a summary of it (Section~\ref{sec:bmc5-metric}).
	A summary like that shows the model the characters mentioned in the passage, its events, their order, and even distinctive words and short exact phrases from the book's actual passage.
	The authors acknowledge this.
	It's why they add a trimming step in the first place not to count contiguous text that overlaps between the prompt and the output.\footnote{``This trimming is necessary because plot summaries often contain exact phrases from the source book''~\citep{liu2026alignmentwhackamolefinetuning}.}
	
	The trouble is, this trimming step only removes contiguous five-word overlaps between an output match and the prompt.
	\textbf{This doesn't remove the very real possibility that a match in a generation comes from reassembling words the summary already supplied.}
	(As with short matches, the paper also doesn't include the right sanity checks to test for false positives here.)
	Handed the right names, events, and order, a capable model can output a span that lines up with the book word-for-word, without ever copying an exact five-word stretch from the prompt.
	
	Consider the following fictitious \textit{Lord of the Rings} fanfic example.
	The (fictitious) ground-truth book contains the text
	
	\begin{quote}
		... That night, Frodo and Sam went down to the marsh ...
	\end{quote}
	
	The prompt (a plot summary of the passage) contains:
	
	\begin{quote}
		... a hobbit named Frodo, together with his companion Sam, decided late one night to go down to the marsh near the Shire ...
	\end{quote}
	
	The model generates:
	
	\begin{quote}
		... It was dark by the time they set out. \textbf{Frodo and Sam went down to the marsh}, keeping to the reeds along the path. ...
	\end{quote}
	
	The part in bold, an 8-word span, matches the (fictitious) book.
	Running \href{https://github.com/cauchy221/Alignment-Whack-a-Mole-Code/blob/fc7dccdbf4efe4863ac28d57716939cf714505dc/evaluation/memorization\_eval\_metrics.py}{the paper's code} on this counts this as a match.
	``Frodo'', ``Sam'', and ``to the marsh'' were all supplied by the prompt, just not as a contiguous 5-word span, so the instruction-trim removes nothing from the count.
	With bmc@5, this gets counted as memorization, even though a high-quality frontier model (like those they run experiments on) could plausibly have produced this output simply by following the instruction to expand that plot summary into a paragraph in the author's style and from the words provided in the prompt.
	(Indeed, I've seen effects exactly like this in my own work for far less capable models.\footnote{I've seen Llama 2 7B models (both chat and base) do something like this in control experiments~\citep{cooper2026principles} to test the validity of a memorization method~\citep{schwarzschild2024rethinking}.})
	My point isn't that the models are \emph{only} ever rearranging the prompt summaries;
	it's that the measurement procedure has no way to tell if a generation reflects this type of rearranging or actual memorization, and it would count both as memorization.\footnote{The paper also reports cross-author fine-tuning results.
		Instead of fine-tuning on the same author for which extraction is attempted for a held-out book, these experiments fine-tune a model on one author's (Haruki Murakami's) books and attempt extraction on held-out books by \emph{other} authors.
		The point is to rule out that fine-tuning merely adapts the model to one author's style, which could confound memorization claims.
		If fine-tuning on Murakami results in generating verbatim text from unrelated authors, the argument is that memorized content must come from training on that text earlier, not the fine-tuning data.
		But this controls only for the fine-tuning author, not for the prompt:
		each held-out book is still probed with a plot summary of its own passages, so a short match can be assembled from that summary, as described in the main text.
		This is the same potential leakage that affects the rest of the coverage numbers.
	}
	
	For GPT-4o in particular, the released pipeline pushes even harder in this direction.
	Although the paper doesn't mention it, the authors give the model a system prompt (at both fine-tuning and inference) that instructs it to use every sentence of the summary, in the order given, and to skip no detail, which turns the task into an order-preserving reconstruction of the summary's content.\footnote{See \href{https://github.com/cauchy221/Alignment-Whack-a-Mole-Code/blob/fc7dccdbf4efe4863ac28d57716939cf714505dc/finetuning/gpt_generate.py\#L57-L68}{\texttt{gpt\_generate.py}} and \href{https://github.com/cauchy221/Alignment-Whack-a-Mole-Code/blob/fc7dccdbf4efe4863ac28d57716939cf714505dc/finetuning/gpt_finetune.py\#L52-L63}{\texttt{gpt\_finetune.py}}. The measurement procedure gives the model 100 chances to produce short matches that would get counted as coverage. This also complicates the paper's cross-model comparison. From what I can tell, the released \href{https://github.com/cauchy221/Alignment-Whack-a-Mole-Code/blob/fc7dccdbf4efe4863ac28d57716939cf714505dc/finetuning/gemini_generate.py\#L53}{Gemini} and \href{https://github.com/cauchy221/Alignment-Whack-a-Mole-Code/blob/fc7dccdbf4efe4863ac28d57716939cf714505dc/finetuning/deepseek_convert.py\#L6-L8}{DeepSeek} generation pipelines don't do this, but the coverage numbers are reported side by side for all three models.}
	
	I'm also \emph{not} suggesting prompt leakage is the mechanism for the long spans produced in single generations.
	If, for instance, a single-generation verbatim span from running the bmc@5 procedure were 200 words, it would be effectively impossible for that output to happen from reassembly.\footnote{This granularity isn't reported for bmc@5.
		As noted in Section~\ref{sec:bmc5}, the paper does report a longest single-generation span --- the ``longest contiguous regurgitated span'' --- but that comes from a different procedure that I'm not discussing here.
		Each generation is matched only against its own excerpt with no instruction trimming, so it isn't a bmc@5 quantity (see the \href{https://github.com/cauchy221/Alignment-Whack-a-Mole-Code/blob/fc7dccdbf4efe4863ac28d57716939cf714505dc/evaluation/memorization_eval_metrics.py\#L417}{released code}).
		The only concrete span length reported for bmc@5 is the longest \emph{merged} block from its aggregated coverage, and, as we've discussed, those merged blocks can be stitched together from very short spans.
	}
	But prompt leakage absolutely is a plausible explanation for the short spans that go into computing bmc@5 results (as discussed in Section~\ref{sec:bmc5-issue1}).
	
	These two problems also compound each other.
	A 5-word span generated verbatim that contains unique words like ``Frodo'' might not seem like a coincidence in the same way that a generic phrase does.
	But if the prompt supplied exactly those unique words, then leakage is still a viable mechanism for that verbatim output.
	
	\subsection{Insufficient experiments to assess false positives}
	\label{sec:insufficient-controls}
	
	Both of the above --- the 5-word minimum and prompt leakage --- are the kind of thing the appropriate control experiment could test, and potentially rule out as problems.
	But the paper doesn't run them.
	
	The relevant type of experiment is called a \textbf{negative control}.
	The idea is to run the same extraction-measurement procedure on data the model couldn't have memorized --- sequences that weren't seen during training.
	Those sequences can't have been memorized, so any extraction signal in this setting would be false positives (Section~\ref{sec:background}).
	Running clean negative controls for extraction is hard, but not impossible --- especially for books.
	One can use books that were published \emph{after} the model's training date.\footnote{I've explored this extensively in my work, including important practical challenges~\citep{cooper2025books, cooper2026principles}.
		A weaker alternative is to change the model, but keep the same data, using a model that wasn't trained on that data~\citep{carlini2023quantifying,cooper2025books}.
		But that is challenging for books and frontier LLMs like those in this work.
	}
	Here, that would've meant running the whole bmc@5 pipeline --- the same fine-tuning, the same plot-summary prompting, the same matching --- on non-training books and reporting the coverage numbers this produced.
	That would've provided a sense of how much of the bmc@5 coverage computed on training books could be due to false positives.
	
	The paper doesn't do this, but it does run other controls.
	For example, they run a synthetic-text experiment:
	instead of fine-tuning on an author's real books, they fine-tune on synthetic stories instead and then run bmc@5.
	The resulting coverage percentages in this setting are lower than for the main experiments on real books.
	The authors claim ``near-zero extraction'' in this setting (even though it's far from zero, I'll get to this below);
	they read this as evidence that the task itself --- expanding a summary in the prompt into prose --- isn't what produces high coverage numbers on training data.
	Instead, they conclude that the underlying mechanism is that the model saw the book during earlier training.\looseness=-1
	
	This is a useful experiment, but it doesn't test the relevant thing that I'm talking about.
	It varies the fine-tuning data, not whether the book being tested was in the original training data.
	As a result, it doesn't validate the paper's headline coverage metric:
	it can show that the task format alone doesn't drive the numbers, but not how much of bmc@5 is due to false positives.
	
	Further, fine-tuning on real books instead of synthetic stories changes the model in a different way, which may confound how much memorization prompting can elicit.
	For instance, fine-tuning on real literary prose (as opposed to synthetic stories) could perhaps make the model better at turning a rich, ordered summary back into real-book-like wording.
	In other words, even though the experiment shows that fine-tuning on real books produces higher coverage than fine-tuning on synthetic stories, it does \emph{not} necessarily show that this higher coverage is actually due to memorization.
	
	Further still, despite the paper's characterization, the results on the synthetic fine-tuning experiments do \emph{not} reflect ``near-zero extraction'' signal using bmc@5.
	It's true that there are ``virtually no long verbatim spans'' for GPT-4o,\footnote{The paper's ``virtually no long verbatim spans'' is a general claim about synthetic fine-tuning, but its own figure shows the other two models do produce long spans on the same book.
		Gemini-2.5-Pro reaches a 64-word single-generation span (26 spans over 20 words), and DeepSeek-V3.1 reaches a 90-word span (11 spans over 20 words).
		Gemini-2.5-Pro has bmc@5 of 28.2\% in this setting, DeepSeek-V3.1 has bmc@5 of 20.3\%.
	} but bmc@5 is high for all three models (their Figure 5).
	For \emph{The Handmaid's Tale} (Margaret Atwood), GPT-4o fine-tuned on synthetic stories has a bmc@5 of 19.4\%.\footnote{The paper also runs another kind of control.
		It measures bmc@5 on the unchanged aligned model, before any fine-tuning.
		The point of this is to isolate how much fine-tuning (whatever data it uses) affects bmc@5.
		For \emph{The Handmaid's Tale}, GPT-4o before fine-tuning scores 6.3\%;
		fine-tuning on synthetic stories more than triples that, to 19.4\%.
		Interestingly, what doesn't move is the longest verbatim span the model produces in any single generation.
		This stays at 18 words.
		This lack of change in single-generation long-span matches is perhaps why the authors call the synthetic fine-tuning results ``near-zero extraction.''
		But, if so, this would also generally cut against the value of the bmc@5 metric in their claims about memorization of significant portions of books.
	}
	Despite the paper's treatment of these different results, it can't be simultaneously true that 19.4\% is ``near-zero extraction'' when interpreting a control and indicative of nearly one-fifth of a whole book being memorized in the main experiments.
	
	\vspace{-.2cm}
	\section{The threat model, cost, and claims about copyright}
	\label{sec:cost}
	\vspace{-.1cm}
	
	It's standard best practice for a paper making security-related claims about production systems to fully specify the threat model.
	An important component is how much the experiments cost, since that's a huge part of what makes this kind of extraction practical or not.
	This is why other work that extracts information from production systems reports cost~\citep{nasr2023scalable,nasr2025scalable,ahmed2026extracting,carlini2024stealing}, but it's glaringly absent here.
	And cost also bears directly on the copyright claims.
	(I mentioned both of these points in my March 2026 email correspondence with the authors, prior to the paper's public release.\footnote{``Cost is important context both for the technical and copyright claims, but I didn’t see this mentioned/discussed in detail.
		We discuss a bit about this in Ahmed et al. 2026, though the considerations here are different (different threat model, significantly increased cost).''
		Email correspondence; March 17, 2026.
	})
	A back-of-the-envelope estimate suggests the full set of experiments in \textit{Alignment Whack-a-Mole} likely cost thousands (if not tens of thousands) of dollars, which has direct implications for the paper's claims about market substitution (see Section~\ref{sec:background}).
    Recall from Section~\ref{sec:bmc5-metric} that running bmc@5 here involved generating 100 times each book's length for every one of the 243 book-model pairs, which corresponds to \textbf{over 24,000 books' worth of generated text}.\footnote{Here is the arithmetic behind what I think is actually a very conservative cost estimate, as the upper end of this range is hit based on back-of-the-envelope calculations for GPT-4o alone.
        Take 100,000 LLM tokens as a conservative stand-in for the length of one book. 
        \emph{Harry Potter and the Sorcerer's Stone} is about 77,000 words, which is a bit over 100,000 tokens at the same 4/3 word-to-token ratio \href{https://github.com/cauchy221/Alignment-Whack-a-Mole-Code/blob/fc7dccdbf4efe4863ac28d57716939cf714505dc/finetuning/gpt_generate.py\#L53}{the paper's own code assumes}.
        (I picked this book because, from my own work, I know its length; see \citet{cooper2025books}.)
        Note also that this isn't a particularly long book; it's about 300 printed pages, which is why it's a conservative stand-in.
        For a single model, 100 generations across 81 books is 8,100 books' worth of output, or roughly 8,100 $\times$ 100,000 = 810 million output tokens.
        The prompts add roughly 600 million input tokens.
        Most of that is the plot summaries: each is \href{https://github.com/cauchy221/Alignment-Whack-a-Mole-Code/blob/fc7dccdbf4efe4863ac28d57716939cf714505dc/preprocess/fix_file.py\#L125}{half the excerpt's length} and is re-sent on every one of the 100 generations, which is about 405 million tokens (half the 810 million output tokens).
        The rest is overhead paid on every call: a one-line instruction wrapper plus two system messages, roughly 110 tokens each time.
        There are roughly 2 million such calls (100 generations for each of the roughly 250 excerpts in a book, across all 81 books), so that overhead adds another 220 million tokens or so.
        At \href{https://developers.openai.com/api/docs/pricing}{OpenAI's list prices} for a fine-tuned GPT-4o (\$15.00 per million output tokens and \$3.75 per million input, as of this writing), that is about \$14,400.
        The released code submits these generations through the \href{https://github.com/cauchy221/Alignment-Whack-a-Mole-Code/blob/fc7dccdbf4efe4863ac28d57716939cf714505dc/finetuning/gpt_generate.py\#L132}{Batch API}, which for this model is discounted to \$12.50 and \$2.225 per million.
        This still comes out to \textbf{about \$11,500 for GPT-4o alone}.
        The fine-tuning itself is a rounding error next to this, on the order of \$300.
        And again, this is only one of the three models, and it's conservative because I've rounded un-reported numbers down at every step (e.g., stand-in book length of 100,000 tokens, 2 million calls for generations based on the low estimate of 250 excerpts in a book).\looseness=-1
    }\looseness=-1
	
	Market substitution, and the ensuing possibility of market harm, is an important and timely question in copyright litigation about generative AI.
	The U.S. Copyright Office's report on generative AI training raised this point directly~\citep{copyright2025ai}.
	In \textit{Kadrey v. Meta}~\citep{kadrey2025meta}, Judge Chhabria explicitly noted that this issue (if demonstrated effectively) could speak to market dilution harm, and therefore cut against fair use.
	(He said this was not done so in that case.)
	The paper is direct about the stakes of this, and implies that its results are directly relevant.
	The authors suggest that fine-tuning lets users ``extract substantial portions of the source works,'' and do so ``with little effort.''
	They argue that the ``regurgitations'' they produce are ``verbatim, or highly similar, copies that could well substitute for the source works.''
	And they ask, ``why comply with a paywall, when one can prompt an AI system to deliver the content unencumbered by access or use restrictions?''\looseness=-1
	
	But, having stepped through exactly what the experiments actually do, the results in \textit{Alignment Whack-a-Mole} clearly don't support this.
	If the headline coverage number of 85--90\% meant that they'd extracted most of a book contiguously, maybe that'd be a different matter.
	But that's not what bmc@5 means.
	Even if we were to assume that (however unlikely) there were no false positives in these results, this 85--90\% comes from pooling across snippets of text;
	those snippets are interwoven with \textit{other} text that isn't the ground-truth book and are isolated from that other text by checking against the actual book.
	They're also potentially out of order, and reassembling them in the correct sequence requires the book itself.
	And the prompts required to produce these snippets also require the actual book, since the summaries they contain are built from ground-truth excerpts.
	All told, a ``covered'' book is a scattered, out-of-order patchwork that requires a copy of the actual book to make sense of (see Figure~\ref{fig:bmc-example}).\looseness=-1
	
	Unlike what the paper suggests, that's nothing like reproducing a whole book with a single initial minimal prompt~\citep{cooper2025books,ahmed2026extracting}, or even a large contiguous stretch of one.\footnote{The paper's largest single-generation verbatim span is around 460 words. That's quite long, but not close to a whole book.\looseness=-1}
	I'm not sure how one can reasonably call this evidence for market substitution, especially when accounting for cost.
	I don't think the question raised here is, ``why comply with a paywall, when one can prompt an AI system to deliver the content unencumbered by access or use restrictions?''
	It's more like, ``why spend hundreds of dollars to obtain scattered fragments of a book that one can buy intact for much less money?''
	
	\vspace{-.2cm}
	\section{Some closing thoughts about the field}
	\label{sec:conclusion}
	\vspace{-.1cm}
	
	I'm now going to take a step back and say some general things, which aren't directly about \textit{Alignment Whack-a-Mole}.
	But I think they're relevant to the broader ecosystem this paper sits in, and to my future in it.\looseness=-1
	
	Working in two different fields is really hard.
	People with different expertise often talk past each other.
	I've written about this a lot with colleagues in both machine learning and law~\citep{cooper2023report,lee2023talkin,cooper2024files,cooper2025unlearning,lemley2026copies}, and done a ton of work to try to bring conceptual clarity to the very complex issues at the intersection of generative AI and law.
	In general, I think the interplay between AI and (the rule of) law is one of the most important topics of our time, and it's something that I also think I'm able to positively contribute to understanding better.
	This is how I originally started working on copyright and AI back in 2020.\looseness=-1
	
	\begin{figure}[t]
		\centering
		\includegraphics[width=0.6\linewidth]{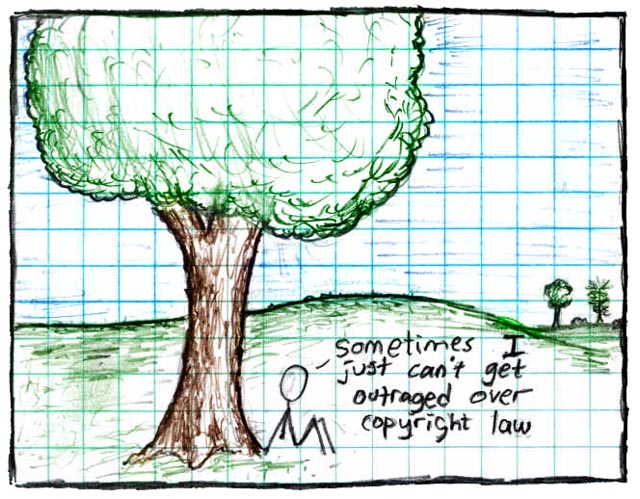}
		\vspace{-.2cm}
		\caption{``Sometimes I just can't get outraged over copyright law.'' Comic by Randall Munroe~\citep{xkcd14}, used under \cclink.}
		\label{fig:xkcd}
		\vspace{-.2cm}
	\end{figure}
	
	This area has gone from a bit of a backwater to front-page news.
	That's been a good thing in many ways.
	It's really exciting to have a group of amazing collaborators.
	I feel tremendously privileged to have long-running dialogues with some of the best legal and AI/ML minds in the world about these issues, let alone to get to share a byline with them.
	This has been the richest and most enjoyable part of my career.
	It's also frankly been an unreal experience to have helped move the needle on lawyers' technical understanding of different parts of machine learning, especially memorization and extraction.
	(Yes, models can be copies of the training data they've memorized.
	But it's complicated, and I don't know if or how much this will matter when it comes to courts making rulings on infringement.)\looseness=-1
	
	But in the last year or so, I've felt a major vibe shift in the ecosystem at large.
	It's not new that everyone on all sides of the copyright debate has strong opinions about it, and that people are generally very angry.
	Something somewhat new, though, is how entangled the field has become with active, high-stakes copyright litigation, and the money and incentives that come with that.
	Don't get me wrong, I think it's genuinely valuable for AI/ML experts to be involved in these cases;
	courts need people who can explain the technical facts.
	But this also means that a growing share of the work in this space comes from people with a personal stake in those cases.
	This is partly why, as a general norm, I think disclosing where funding comes from and \href{https://researchconnect.stonybrook.edu/en/projects/mosaic-large-language-models/}{potential conflicts of interest} is a bare minimum requirement, as a matter of research ethics.
	(I think this is true in general for science, but especially in this field.)\looseness=-1
	
	For the last several years working in this area, I've tried really hard to focus on putting out careful work that presents evidence others can rely on.
	But this is becoming much harder to do.
	Many recent papers on copyright and AI rest on fundamental misconceptions about memorization and extraction, so much so that it feels like I'm spending most of my time playing Whack-a-Mole:
	for each one I clear up, five more crop up in its place.
	A lot of these misconceptions are due to honest mistakes, but some come from extreme sloppiness (human and AI alike) or an interest in a certain argument winning the day (regardless of the validity of the evidence on which that argument is built).\looseness=-1
	
	And in this landscape, I'm not having a ton of fun anymore.
	Writing something like this isn't fun.
	It also takes an enormous amount of time that could have been spent doing literally anything else.
	So I'll be taking a step back for now from starting new work in this area, and will be turning my attention to other areas where I can hopefully make useful contributions and have some fun along the way.
	In the meantime, I'm still available over email to answer questions about memorization, extraction, and copyright.
	(But I get a lot of them, so please forgive me if I'm slow in responding.)
	
	\bibliographystyle{plainnat}
	\bibliography{refs}
	
\end{document}